Diquarks and pentaquark nature.

S.M.Gerasyuta[1,2], V.I.Kochkin[1]

Department of Theoretical Physics St.Petersburg University
198904, St.Petersburg, Russia[1]

Department of Physics, LTA, 194021 St.Petersburg, Russia[2]

Abstract

The relativistic five-quark equations are found in the framework of the dispersion relation technique. The five-quark amplitudes for the low-lying pentaquarks including the u,d,s-quarks are calculated. The poles of these amplitudes determine the masses of N- and Xi- pentaquarks. The mass spectra of the lowest pentaquarks with the $J^P = 1/2^+, 1/2^-, 3/2^+$ are calculated. The mass values of the positive and negative parity pentaquarks are determined by the mixing of both $0^+$ and $1^+$ diquarks.

## I. Introduction

The existence of particles made of more than three quarks is an important issue of QCD inspired models. Recently several experimental groups have reported observation of a new exotic (S=1) baryon resonance $\Theta^+$(1540) [1-10] with a narrow decay width. Then the NA49 group at CERN announced evidence for an additional narrow "cascade" exotic (Q=S=-2) $\Xi^{--}$ with mass close to 1860 [11, 12].

The chiral soliton or Skyrme model has played an important role in stimulating the search for exotic [13-18]. Diakonov et al. [19] have studied the properties of $\Theta^+$ as a member of antidecuplet with prediction of mass and width are close to the results of experimentalist.

Since the $\Theta^+$ discovery, there has been a flurry of papers studying pentaquarks in the constituent quark model [20-22], correlated quark model [23, 24], chiral soliton papers [25-27].

The spin and parity quantum numbers of the $\Theta^+$ have yet to be determined experimentally. The spin of $\Theta^+$ is taken to be 1/2 by all theory papers and various estimates show that spin-3/2 pentaquarks must be heavier [28-30]. A more controversal point among theorist is the parity of the state. For example, QCD sum rule calculation [31, 32], quenched lattice QCD [33, 34] predict that the light $\Theta^+$ is a negative parity isosinglet. All other papers [20-30] predict the lightest $\Theta^+$ pentaquark as a positive parity state.

The photoproduction and pion-induced production cross sections of the $\Theta^+$ were studied in [35]. It is shown in both cases that the production cross sections for a negative parity $\Theta^+$ are much smaller than those for the positive parity state (for a given $\Theta^+$ width). In Ref. [35], results for the $\Theta^+$ production cross section in photon-proton reactions were compared with estimates of the cross section based on data obtained by the SAPHIR Collaboration [4], odd-parity pentaquarks were arqued to be disfavored.

Very recently, an unambiguous method $\Theta^+$ production near threshold for a polarized proton reaction was proposed [36, 37]. The cross section for the allowed spin configuration are estimated. It is of order of one microbarn for the positive parity $\Theta^+$ and about 0.1 microbarn for the negative parity $\Theta^+$ in the threshold energy region.

In our papers [38, 39] the generalization of five-quark equations (like Faddeev-Yakubovsky approach) are constructed in the form of the dispersion relation. The five-quark amplitudes for the low-lying pentaquarks contain only u,d-quarks [39]. The poles of these amplitudes determine the masses of the $udud\bar{d}$ amplitude contains only the diquark with $J^P = 0^+$, the $\Delta$-isobar pentaquark amplitude is constructed using the diquark with $J^P = 1^+$. In our relativistic quark model with four-fermion interaction in diquark channel we have the diquark level $D$ with $J^P = 0^+$ and the mass $m_{u,d} = 0.72 GeV$ (in the color state $3_c$) [40]. The diquark state with $J^P = 1^+$ in color state $3_c$ also has an attractive interaction, but smaller than that of the diquark with $J^P = 0^+$, therefore there is only the correlation of quarks. In this case the low-lying $\Delta$-isobar pentaquark masses are smaller than the $N$-pentaquark masses [39]. It depends on the different interactions in the diquark channels $J^P = 0^+, 1^+$.

In our previous paper the relativistic five-quark equations for the family of the $\Theta^+$ pentaquarks are constructed [41]. The five-quark amplitudes for the low-lying $\Theta^+$ pentaquarks are calculated. The poles of these amplitudes determine the masses of the $\Theta^+$ pentaquarks. The masses of the constituent u, d, s -quarks coincide with the quark masses of the ordinary baryons in our quark model [42]: $m_{u,d} = 410 MeV$, $m_s = 557 MeV$.

We received the masses $\Theta^+$ pentaquarks with $J^P = 1/2^+, 1/2^-, 3/2^+, 3/2^-$ and predict the masses of $\Theta^{++}(\Theta^0)$ pentaquarks. The gluon coupling constant $g = 0.456$ is determined by fixing of $\Theta^+$ pentaquark mass with $I_z = 0$ and $J^P = 1/2^+$. The model in consideration has only this new parameter as compared previous paper [39]. The cut-off parameters coincide with those in [39]: $\Lambda_{0^+} = 16.5$ and $\Lambda_{1^+} = 20.12$ for the diquark with $0^+$ and $1^+$ respectively. The cut-off parameter for the mesons is equal to 16.5.

The mass of $\Theta$ pentaquark with positive parity are obtained smaller than the mass of $\Theta$ pentaquark with negative parity. It depends on the mixing of $0^+$ and $1^+$ diquarks.

Our calculations take into account the contributions to the $\Theta^+$ amplitude not only 8+10*- plets but also 35-, 27-, 10- plets of $SU(3)_f$.

The present paper is devoted to the construction of relativistic five-quark equations for the nucleon and $\Xi^{--}$ pentaquarks. The five-quark amplitudes for lowest pentaquarks (nucleon family) contain only u, d- quarks. The poles of these amplitudes determine the masses of the $udud\bar{u}$ pentaquarks. We considered two cases:

1. using only the diquark $0^+$ (Table1) and
2. including the $0^+$ and $1^+$ diquarks (Table 2).

In the first case the negative parity state with $J^P = 1/2^+$ is lower than positive parity state. In the second case we construct the nucleon using the $0^+$ and $1^+$ diquarks and receive the positive parity nucleon state lighter than negative. This nucleon state with the mass M=1480MeV can be considered as the Roper resonance. The important result of this model is the mixing of diquark $0^+$ and $1^+$. The mass values of the low-lying nucleon pentaquarks are shown in the Tables 1and 2.

The masses of $\Xi^{--}$ family was also calculated (Table3). The mass of positive parity $\Xi^{--}$ state with the $J^P = 1/2^+$ and $I = 3/2$ is smaller than the mass of state with negative parity.

In this case the important role also plays the mixing $0^+$ and $1^+$ diquarks. Our calculation take into account the contribution to the $\Xi^{--}$ amplitude the 35-, 27-, 10-, 10*-, 8- plets.

The paper is organized as follows. After this introduction, we discuss the five-quark amplitudes which contain u, d, s -quarks (Section II). In the section III, we report our numeral results (Tables 1, 2, 3) and the last section is devoted to our discussion and conclusion.

## II. Pentaquark amplitudes.

We derived the relativistic five-quark equations in the framework of the dispersion relation technique. We use only planar diagrams; the other diagrams due to the rules of $1/N_c$ expansion [43-45] are neglected. The correct equations for the amplitude are obtained by taking into account all possible subamplitudes. It corresponds to the division of the complete system into subsystems with a smaller number of particles. Then one should represent a five-particle amplitude as a sum of ten subamplitudes: $A = A_{12} + A_{13} + A_{14} + A_{15} + A_{23} + A_{24} + A_{25} + A_{34} + A_{35} + A_{45}$. We need to consider only one group of diagrams and the amplitude corresponding to them, for example $A_{12}$. The set of diagrams associated with the amplitude $A_{12}$ (the nucleon pentaquarks with the $0^+$ diquark) can be further broken down into four groups corresponding to amplitudes $A_1(s, s_{1234}, s_{12}, s_{34})$, $A_2(s, s_{1234}, s_{25}, s_{34})$, $A_3(s, s_{1234}, s_{13}, s_{134})$, $A_4(s, s_{1234}, s_{23}, s_{234})$ (Fig. 1). The antiquark is shown by the arrow and the other lines correspond to the quarks. The coefficients are determined by the permutation of quarks [46, 47].

In order to represent the subamplitudes $A_1(s, s_{1234}, s_{12}, s_{34})$, $A_2(s, s_{1234}, s_{25}, s_{34})$, $A_3(s, s_{1234}, s_{13}, s_{134})$, and $A_4(s, s_{1234}, s_{23}, s_{234})$ in the form of a dispersion relation it is necessary to define the amplitudes of quark-quark and quark-antiquark interaction $b_n(s_{ik})$. The pair quarks amplitudes $q\bar{q} \to q\bar{q}$ and $qq \to qq$ are calculated in the framework of the dispersion N/D method with the input four-fermion interaction with quantum numbers of the gluon [40]. We use the results of our relativistic quark model [40] and write down the pair quarks amplitude in the form:

$$b_n(s_{ik}) = \frac{G_n^2(s_{ik})}{1 - B_n(s_{ik})}, \quad (1)$$

$$B_n(s_{ik}) = \int_{(m_1+m_2)^2}^{\Lambda_n} \frac{ds'_{ik}}{\pi} \frac{\rho_n(s'_{ik}) G_n^2(s'_{ik})}{s'_{ik} - s_{ik}}. \quad (2)$$

Here $s_{ik}$ is the two-particle subenergy squared, $s_{ijk}$ corresponds to the energy squared of particles $i, j, k$, $s_{ijkl}$ is the four-particle subenergy squared and $s$ is the system total energy squared. $G_n(s_{ik})$ are the quark-quark and quark-antiquark vertex functions (Table 4). $B_n(s_{ik})$, $\rho_n(s_{ik})$ are the Chew-Mandelstam functions with the cut – off $\Lambda_n$ ($\Lambda_1 = \Lambda_3$) [48] and the phase spaces respectively.

The two-particle phase space for the equal quark masses is defined as:

$$\rho_n(s_{ik}, J^{PC}) = \left(\alpha(J^{PC}, n) \frac{s_{ik}}{(m_i + m_k)^2} + \beta(J^{PC}, n)\right) \frac{\sqrt{[s_{ik} - (m_i + m_k)^2][s_{ik} - (m_i - m_k)^2]}}{s_{ik}},$$

The coefficients $\alpha(J^{PC}, n)$ and $\beta(J^{PC}, n)$ are given in Table 5.
Here n=1 corresponds to a $qq$-pair with $J^P = 0^+$ in the $\bar{3}_c$ color state, n=2 describes a $qq$-pair with $J^P = 1^+$ in the $\bar{3}_c$ color state and n=3 defines the $q\bar{q}$-pairs, corresponding to mesons with quantum numbers: $J^{PC} = 0^{++}, 0^{-+}$ and $I_z = 0,1$.

In the case in question the interacting quarks do not produce a bound state; therefore the integration in Eqs.(3) - (6) below is carried out from the threshold $(m_i + m_k)^2$ to the cut-off $\Lambda_n$. The system of integral equations, corresponding to Fig. 1 (the meson state with $J^{PC} = 0^{++}$ and diquark with $J^P = 0^+$), can be described as:

$$A_1(s, s_{1234}, s_{12}, s_{34}) = \frac{\lambda_1 B_3(s_{12}) B_1(s_{34})}{[1 - B_3(s_{12})][1 - B_1(s_{34})]} + 3\hat{J}_2(3,1) A_4(s, s_{1234}, s'_{23}, s'_{234}) +$$
$$+ 2\hat{J}_2(3,1) A_3(s, s_{1234}, s'_{13}, s'_{134}) + 2\hat{J}_1(3) A_3(s, s_{1234}, s'_{15}, s'_{125}) + \qquad , \quad (3)$$
$$+ 2\hat{J}_1(3) A_4(s, s_{1234}, s'_{25}, s'_{125}) + 2\hat{J}_1(1) A_4(s, s_{1234}, s'_{35}, s'_{345})$$

$$A_2(s,s_{1234},s_{25},s_{34}) = \frac{\lambda_2 B_1(s_{25})B_1(s_{34})}{[1-B_1(s_{25})][1-B_1(s_{34})]} + 6\hat{J}_2(1,1)A_4(s,s_{1234},s'_{23},s'_{234}) + \qquad (4)$$

$$+ 8\hat{J}_1(1)A_3(s,s_{1234},s'_{12},s_{125})$$

$$A_3(s,s_{1234},s_{13},s_{134}) = \frac{\lambda_3 B_3(s_{13})}{1-B_3(s_{13})} + 8\hat{J}_3(3)A_1(s,s_{1234},s'_{12},s'_{34}), \qquad (5)$$

$$A_4(s,s_{1234},s_{23},s_{234}) = \frac{\lambda_4 B_1(s_{23})}{1-B_1(s_{23})} + 2\hat{J}_3(1)A_2(s,s_{1234},s'_{25},s'_{34}) + \qquad (6)$$

$$+ 2\hat{J}_3(1)A_1(s,s_{1234},s'_{12},s'_{34})$$

were $\lambda_i$ are the current constants. We introduced the integral operators:

$$\hat{J}_1(l) = \frac{G_l(s_{12})}{[1-B_l(s_{12})]} \int_{(m_i+m_k)^2}^{\Lambda_n} \frac{ds'_{12}}{\pi} \frac{G_l(s'_{12})\rho_l(s'_{12})}{s'_{12}-s_{12}} \int_{-1}^{+1} \frac{dz_1}{2}, \qquad (7)$$

$$\hat{J}_2(l,p) = \frac{G_l(s_{12})G_p(s_{34})}{[1-B_l(s_{12})][1-B_p(s_{34})]} \times$$

$$\times \int_{(m_i+m_k)^2}^{\Lambda_n} \frac{ds'_{12}}{\pi} \frac{G_l(s'_{12})\rho_l(s'_{12})}{s'_{12}-s_{12}} \int_{(m_i+m_k)^2}^{\Lambda_n} \frac{ds'_{34}}{\pi} \frac{G_p(s'_{34})\rho_p(s'_{34})}{s'_{34}-s_{34}} \int_{-1}^{+1} \frac{dz_3}{2} \int_{-1}^{+1} \frac{dz_4}{2}, \qquad (8)$$

$$\hat{J}_3(l) = \frac{G_l(s_{12},\tilde{\Lambda})}{1-B_l(s_{12},\tilde{\Lambda})} \times$$

$$\times \frac{1}{4\pi} \int_{(m_i+m_k)^2}^{\tilde{\Lambda}} \frac{ds'_{12}}{\pi} \frac{G_l(s'_{12},\tilde{\Lambda})\rho_l(s'_{12})}{s'_{12}-s_{12}} \int_{-1}^{+1} \frac{dz_1}{2} \int_{-1}^{+1} dz \int_{z_2^-}^{z_2^+} dz_2 \frac{1}{\sqrt{1-z^2-z_1^2-z_2^2+2zz_1z_2}}, \qquad (9)$$

were $l, p$ are equal 1 or 3. If we use the diquark state with $J^P = 1^+$ and the meson with $J^{PC} = 0^{++}, 0^{-+}$, $l, p$ are equal 2 or 3. Here $m_i$ is a quark mass.

In Eqs.(7) and (9) $z_1$ is the cosine of the angle between the relative momentum of the particles 1 and 2 in the intermediate state and the momentum of the particle 3 in the final state, taken in the c.m. of particles 1 and 2. In Eq.(9) $z$ is the cosine of the angle between the momenta of the particles 3 and 4 in the final state, taken in the c.m. of particles 1 and 2. $z_2$ is the cosine of the angle between the relative momentum of particles 1 and 2 in the intermediate state and the momentum of the particle 4 in the final state, taken in the c.m. of particles 1 and 2. In Eq. (8): $z_3$ is the cosine of the angle between relative momentum of particles 1 and 2 in the intermediate state and the relative momentum of particles 3 and 4 in the intermediate state, taken in the c.m. of particles 1 and 2. $z_4$ is the cosine of the angle between the relative momentum of the particles 3 and 4 in the intermediate state and momentum of the particle 1 in the intermediate state, taken in the c.m. of particles 3, 4.

We can pass from the integration over the cosines of the angles to the integration over the subenergies [49].

Let us extract two-particle singularities in the amplitudes $A_1(s,s_{1234},s_{12},s_{34})$, $A_2(s,s_{1234},s_{25},s_{34})$, $A_3(s,s_{1234},s_{13},s_{134})$, and $A_4(s,s_{1234},s_{23},s_{234})$:

$$A_1(s,s_{1234},s_{12},s_{34}) = \frac{\alpha_1(s,s_{1234},s_{12},s_{34})B_3(s_{12})B_1(s_{34})}{[1-B_3(s_{12})][1-B_1(s_{34})]}. \qquad (10)$$

$$A_2(s,s_{1234},s_{25},s_{34}) = \frac{\alpha_2(s,s_{1234},s_{25},s_{34})B_1(s_{25})B_1(s_{34})}{[1-B_1(s_{25})][1-B_1(s_{34})]}. \qquad (11)$$

$$A_3(s,s_{1234},s_{13},s_{134}) = \frac{\alpha_3(s,s_{1234},s_{13},s_{134})B_3(s_{13})}{1-B_3(s_{13})}, \tag{12}$$

$$A_4(s,s_{1234},s_{23},s_{234}) = \frac{\alpha_4(s,s_{1234},s_{23},s_{234})B_1(s_{23})}{1-B_1(s_{23})}, \tag{13}$$

We do not extract three- and four-particle singularities, because they are weaker than two-particle singularities.

We used the classification of singularities, which was proposed in paper [50] for the two and three particle singularities. The construction of approximate solution of Eqs.(3) - (6) is based on the extraction of the leading singularities of the amplitudes. The main singularities in $s_{ik} \approx (m_i + m_k)^2$ are from pair rescattering of the particles i and k. First of all there are threshold square-root singularities. Also possible singularities are pole singularities which correspond to the bound states. The diagrams of Fig.1 apart from two-particle singularities have triangular singularities and singularities defining the interaction of four and five particles. Such classification allowed us to find the corresponding solutions of Eqs.(3) - (6) by taking into account some definite number of leading singularities and neglecting all the weaker ones. We considered the approximation which defines two-particle, triangle, four- and five-particle singularities. The functions $\alpha_1(s,s_{1234},s_{12},s_{34})$, $\alpha_2(s,s_{1234},s_{25},s_{34})$, $\alpha_3(s,s_{1234},s_{13},s_{134})$, and $\alpha_4(s,s_{1234},s_{23},s_{234})$ are smooth functions of $s_{ik}$, $s_{ijk}$, $s_{ijkl}$, $s$ as compared with the singular part of the amplitudes, hence they can be expanded in a series in the singularity point and only the first term of this series should be employed further. Using this classification one define the reduced amplitudes $\alpha_1$, $\alpha_2$, $\alpha_3$, $\alpha_4$ as well as the B-functions in the middle point of the physical region of Dalitz-plot at the point $s_0$:

$$s_0^{ik} = s_0 = \frac{s+3\sum_{i=1}^{5}m_i^2}{0.25\sum_{\substack{i,k=1 \\ i\neq k}}^{5}(m_i+m_k)^2}, \tag{14}$$

$$s_{123} = 0.25 s_0 \sum_{\substack{i,k=1 \\ i\neq k}}^{3}(m_i+m_k)^2 - \sum_{i=1}^{3}m_i^2, \quad s_{1234} = 0.25 s_0 \sum_{\substack{i,k=1 \\ i\neq k}}^{4}(m_i+m_k)^2 - 2\sum_{i=1}^{4}m_i^2$$

Such a choice of point $s_0$ allows to replace the integral equations (3) - (6) (Fig.1) by the algebraic equations (15) - (18) respectively:

$$\alpha_1 = \lambda_1 + 3\alpha_4 J_2(3,1,1) + 2\alpha_3 J_2(3,1,3) + 2\alpha_3 J_1(3,3) + 2\alpha_4 J_1(3,1) + 2\alpha_4 J_1(1,1), \tag{15}$$

$$\alpha_2 = \lambda_2 + 6\alpha_4 J_2(1,1,1) + 8\alpha_3 J_1(1,3), \tag{16}$$

$$\alpha_3 = \lambda_3 + 8\alpha_1 J_3(3,3,1), \tag{17}$$

$$\alpha_4 = \lambda_4 + 2\alpha_2 J_3(1,1,1) + 2\alpha_1 J_3(1,1,3). \tag{18}$$

We use the functions $J_1(l,p)$, $J_2(l,p,r)$, $J_3(l,p,r)$ ($l,p,r = 1, 2, 3$):

$$J_1(l,p) = \frac{G_l^2(s_0^{12})B_p(s_0^{15})}{B_l(s_0^{12})} \int_{(m_i+m_k)^2}^{\Lambda_n} \frac{ds'_{12}}{\pi} \frac{\rho_l(s'_{12})}{s'_{12}-s_0^{12}} \int_{-1}^{+1}\frac{dz_1}{2}\frac{1}{1-B_p(s'_{15})}, \tag{19}$$

$$J_2(l,p,r) = \frac{G_l^2(s_0^{12})G_p^2(s_0^{34})B_r(s_0^{13})}{B_l(s_0^{12})B_p(s_0^{34})} \times$$

$$\times \int_{(m_i+m_k)^2}^{\Lambda_n} \frac{ds'_{12}}{\pi} \frac{\rho_l(s'_{12})}{s'_{12}-s_0^{12}} \int_{(m_i+m_k)^2}^{\Lambda_n} \frac{ds'_{34}}{\pi} \frac{\rho_p(s'_{34})}{s'_{34}-s_0^{34}} \int_{-1}^{+1} \frac{dz_3}{2} \int_{-1}^{+1} \frac{dz_4}{2} \frac{1}{1-B_r(s'_{13})}$$

(20)

$$J_3(l,p,r) = \frac{G_l^2(s_0^{12},\tilde{\Lambda})B_p(s_0^{13})B_r(s_0^{24})}{1-B_l(s_0^{12},\tilde{\Lambda})} \frac{1-B_l(s_0^{12})}{B_l(s_0^{12})} \frac{1}{4\pi} \times$$

$$\times \int_{(m_i+m_k)^2}^{\tilde{\Lambda}} \frac{ds'_{12}}{\pi} \frac{\rho_l(s'_{12})}{s'_{12}-s_0^{12}} \int_{-1}^{+1} \frac{dz_1}{2} \int_{-1}^{+1} dz \int_{z_2^-}^{z_2^+} dz_2 \frac{1}{\sqrt{1-z^2-z_1^2-z_2^2+2zz_1z_2}} \frac{1}{[1-B_p(s'_{13})][1-B_r(s'_{24})]}$$

(21)

The other choices of point $s_0$ do not change essentially the contributions of $\alpha_1$, $\alpha_2$, $\alpha_3$ and $\alpha_4$; therefore we omit the indexes $s_0^{ik}$. Since the vertex functions depend only slightly on energy it is possible to treat them as constants in our approximation and determine them in a way similar to that used in [49]. We describe the integration contours of functions $J_1, J_2, J_3$ similar to the paper [49]. These contours are determined by the interaction of the quarks.

The solutions of the system of equations are considered as:

$$\alpha_i(s) = F_i(s,\lambda_i)/D(s) \qquad (22)$$

where zeros of $D(s)$ determinants define the masses of bound states of pentaquarks. $F_i(s,\lambda_i)$ are the functions of $s$ and $\lambda_i$. The functions $F_i(s,\lambda_i)$ determine the contributions of subamplitudes to the pentaquark amplitude.

### III. Calculation results

For the sake of simplicity we considered in detail the calculation of the nucleon pentaquark masses only using the diquark $0^+$ (Fig.1). The calculations of the masses of nucleon pentaquarks with the diquarks $0^+$ and $1^+$ (Fig.2) and the $\Xi^{--}$ pentaquark masses (Fig.3) are similar to the first case but more unwieldy.

The poles of the reduced amplitudes correspond to the bound states and determine the masses of nucleon (Fig.1, 2) and $\Xi^{--}$ (Fiq.3) pentaquarks. The quark masses of model $m_{u,d} = 410 MeV$ and $m_s = 557 MeV$ coincide with the quark masses of ordinary baryons in our model [42]. The model in consideration has not new parameter as compared previous papers[39, 41]. The gluon coupling constant $g = 0.456$ is determined by fixing mass 1540 MeV. The cut-off parameters coincide with those in paper [39] for the diquark with $0^+$ and $1^+$ :16.5 and 20.12 respectively. The cut-off parameter for the mesons is equal to 16.5. The calculated mass values of the nucleon pentaquarks are shown in the Table1 (only with $0^+$ diquarks). In the Table 2 one take into account both $0^+$ and $1^+$ diquarks. The masses of nucleon pentaquarks ($udud\bar{u}$) with $J^P = 1/2^+, 1/2^-, 3/2^+$ are predicted. The masses of $\Xi^{--}$ ($ssdd\bar{u}$) with S= Q= -2 are calculated. The masses of $\Xi^{--}$ state with the positive parity is smaller than the masses of state with negative parity. The important role play the mixing of $0^+$ and $1^+$ diquarks (Fig.3).

## IV. Conclution

In our relativistic five-quark model (Faddeev-Yakubovsky type) the masses of low-lying pentaquark are calculated. We used the u, d, s -quarks. The quark amplitudes obey the global color symmetry and include the contributions of the 35-, 27-, 10*-, 10-, 8- plets. The masses of the constituent quarks are equal $m_{u,d} = 410 MeV$ and $m_s = 557 MeV$.

We considered the scattering amplitudes of the constituent quarks. The poles of these amplitudes determine the masses of low-lying pentaquarks. The derived five-quark amplitudes consist of the subamplitudes Fig.1, 2, 3. The nucleon pentaquark with $J^P = 1/2^+$ and mass M=1480MeV can be considered as the Roper resonance.

Unlike meson, all half-integral spin and parity quantum numbers are allowed in baryon sector, so that experiments search for nucleon pentaquark states are not simple. Furthemore, no decay channels are a forbidden. These two facts make identification of a pentaquark difficult.

The exotic $\Xi^{--}$ (S= -2) state has not the three quarks nature and it is natural to interpret as the pentaquark.

We have been treating the quarks as real particles. However, in the soft region the quark diagrams should be treated as spectral integral of the quark mass with spectral density $\rho(m^2)$: the inteqration of the quark mass in the amplitudes and introduces the hadron ones. One can believe that the approximation:

$$\rho(m^2) \Rightarrow \delta(m^2 - m_q^2) \qquad (23)$$

Could be possible for the low-lying hadrons (here $m_q$ is the mass of the constituent quark). We hope that approach given by (23) sufficiently good for the calculation of the low-lying pentaquarks carried at here.

The interesting research is the consideration of $udud\bar{c}$ states. Their decay to N($qQ$) also violates isospin consideration. It is similar to the case of $\Theta^+$.

## V. Acknowledgment

One of the authors (S.M.Gerasyuta) is indebted to the RCNP of Osaka University for hospitality. The authors thank the Yukawa Institute for Theoretical Physics at Kyoto University, where this work was completed during the YITP-W-03-21 on "Multi-quark Hadron: four, five and more?" The authors would like to thank T.Barnes, D.Diakonov, A.Hosaka, B.Silvestre-Brac, H.Toki.8

Table I. Low-lying nucleon pentaquark masses (diquark with $J^P = 0^+$).

| Meson $J^{PC}$ | $J^P$ | Mass, MeV |
|---|---|---|
| $0^{++}$ | $\frac{1}{2}^+$ | 1686 |
| $0^{-+}$ | $\frac{1}{2}^-$ | 1583 |

Parameters of model: quark mass $m_{u,d} = 410$ MeV, cut-off parameter $\Lambda_{0^+} = 16.5$; gluon coupling constant $g = 0.417$.

Table II. Low-lying nucleon pentaquark masses (diquark with $J^P = 0^+$ and $J^P = 1^+$).

| Meson $J^{PC}$ | $J^P$ | Mass, MeV |
|---|---|---|
| $0^{++}$ | $\frac{1}{2}^+$ | 1480 |
| $0^{++}$ | $\frac{3}{2}^+$ | 1900 |
| $0^{-+}$ | $\frac{1}{2}^-$ | 1515 |

Parameters of model: quark mass $m_{u,d} = 410$ MeV, cut-off parameter $\Lambda_{0^+} = 16.5$, $\Lambda_{1^+} = 20.12$; gluon coupling constant $g = 0.417$.

Table III. Low-lying $\Xi$ - pentaquark masses (diquark with $J^P = 0^+$ and $J^P = 1^+$).

| Meson $J^{PC}$ | $J^P$ | Mass, MeV |
|---|---|---|
| $0^{++}$ | $\frac{1}{2}^+$ | 1673 |
| $0^{++}$ | $\frac{3}{2}^+$ | 2200 |
| $0^{-+}$ | $\frac{1}{2}^-$ | 1936 |

Parameters of model: quark mass $m_{u,d} = 410$ MeV, $m_s = 557$ MeV; cut-off parameter $\Lambda_{0^+} = 16.5$, $\Lambda_{1^+} = 20.12$; gluon coupling constant $g = 0.456$.

Table IV. Vertex functions

| $J^{PC}$ | $G_n^2$ |
|---|---|
| $0^+$ (n=1) | $4g/3 - 8gm^2/(3s_{ik})$ |
| $1^+$ (n=2) | $2g/3$ |
| $0^{++}$ (n=3) | $8g/3$ |
| $0^{-+}$ (n=3) | $8g/3 - 16gm^2/(3s_{ik})$ |

Table V. Coefficient of Chew-Mandelstam functions
for n = 3 (meson states) and diquarks n = 1 ($J^P = 0^+$), n = 2 ($J^P = 1^+$).

| $J^{PC}$ | n | $\alpha(J^{PC},n)$ | $\beta(J^{PC},n)$ |
|---|---|---|---|
| $0^+$ | 1 | 1/2 | 0 |
| $1^+$ | 2 | 1/3 | 1/6 |
| $0^{++}$ | 3 | 1/2 | -1/2 |
| $0^{-+}$ | 3 | 1/2 | 0 |

Figure captions.

Fig.1. Graphic representation of the low-lying nucleon pentaquark equations (diquark with $J^P = 0^+$).

Fig.2. Low-lying nucleon pentaquarks (diquarks with $J^P = 0^+$ and $J^P = 1^+$).

Fig.3. Low-lying $\Xi^{--}$- pentaquarks (diquarks with $J^P = 0^+$ and $J^P = 1^+$).

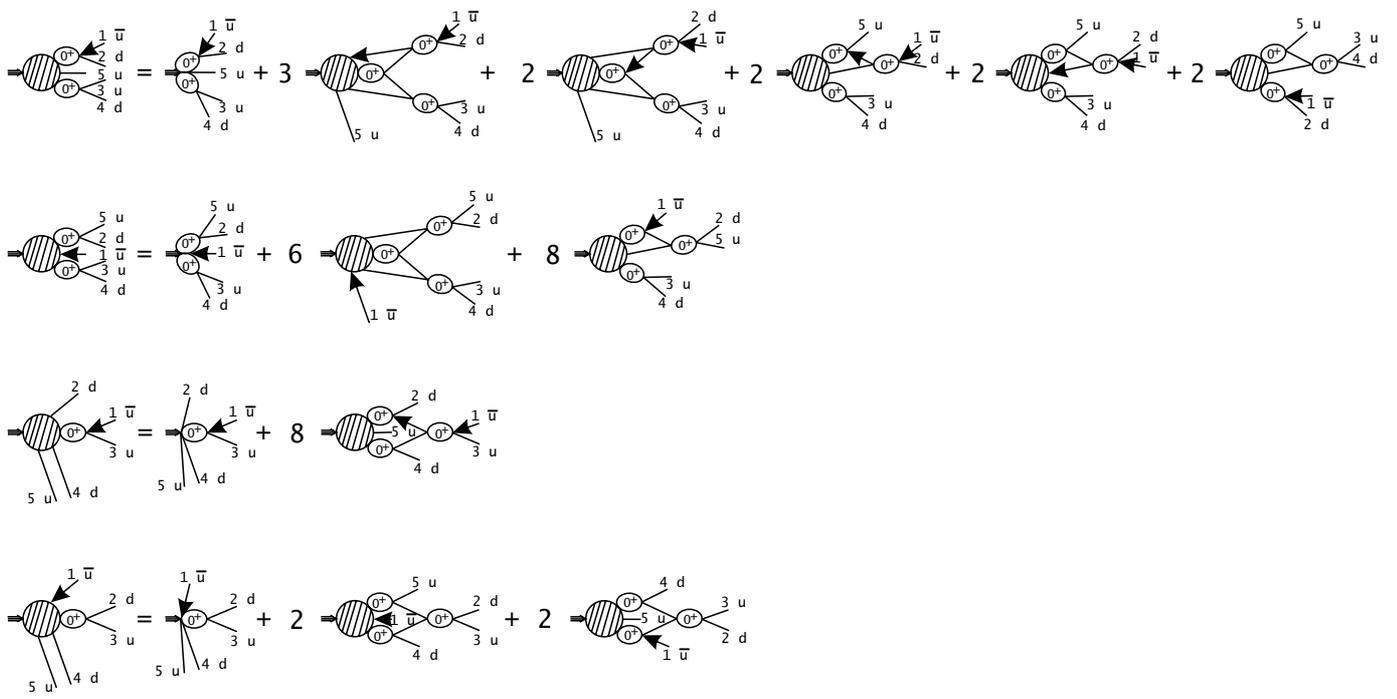

Fig.1

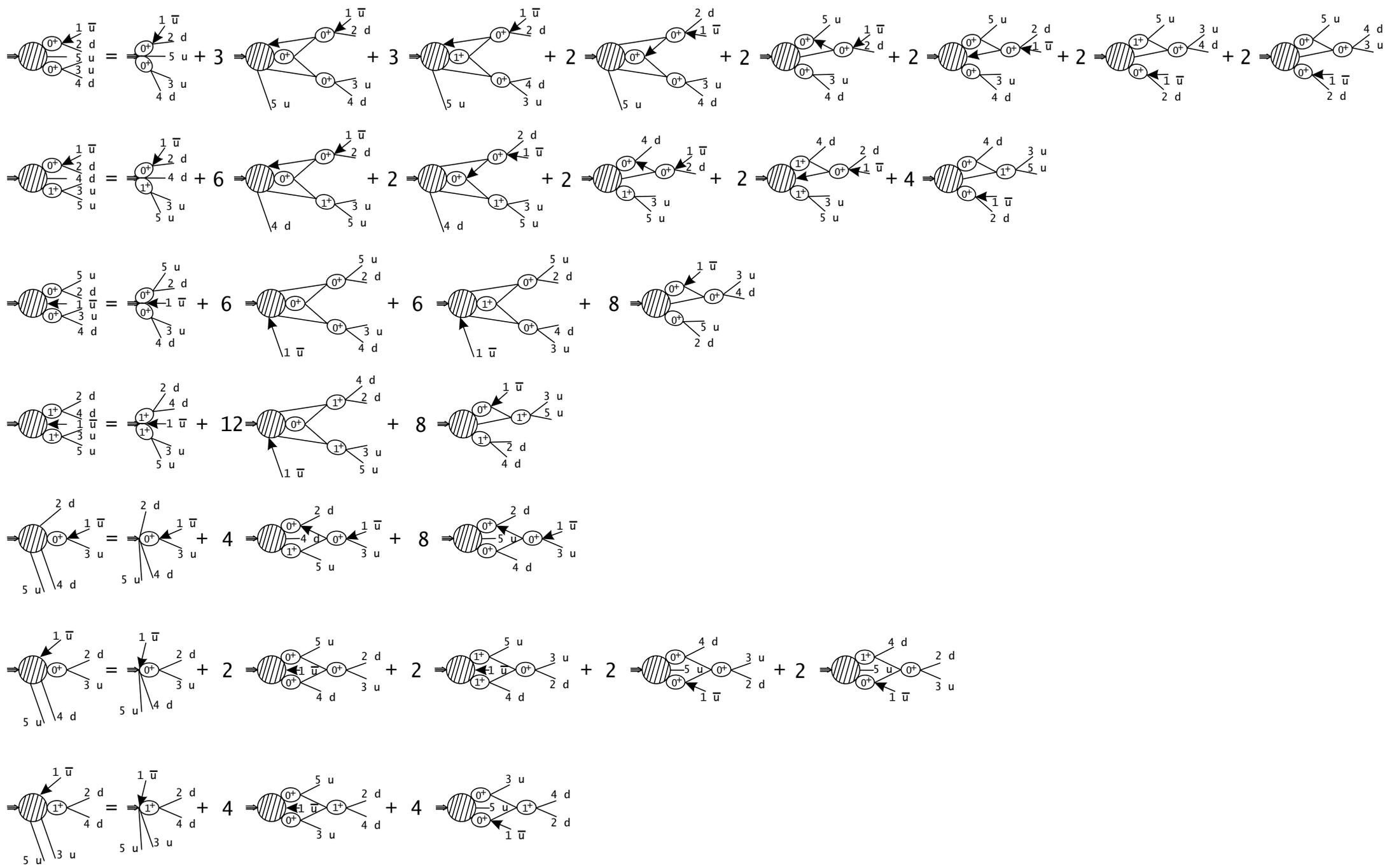

Fig.2

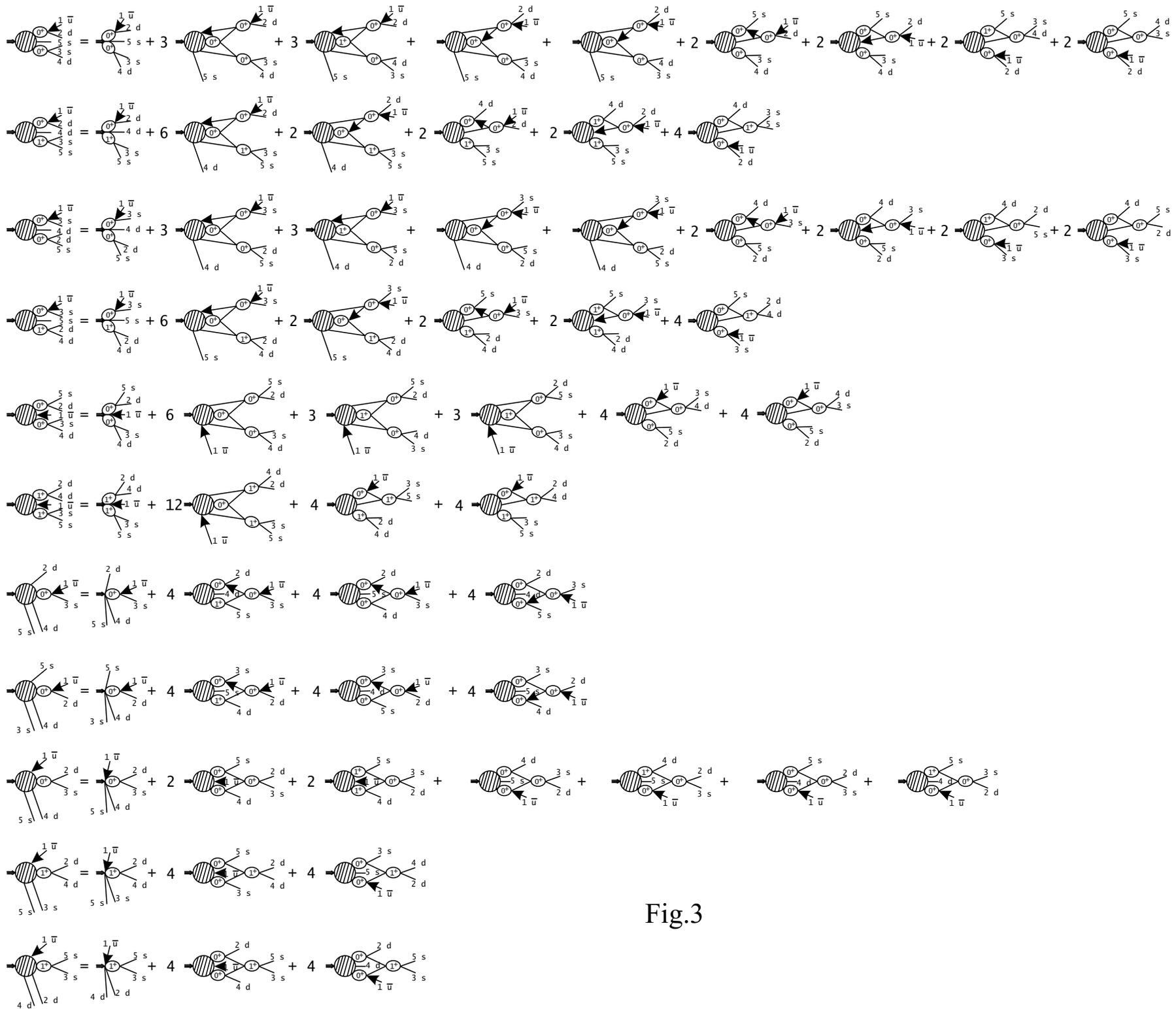

Fig.3